\documentclass{article}
\usepackage{arxiv}

\usepackage[utf8]{inputenc} 
\usepackage[T1]{fontenc}    
\usepackage{hyperref}       
\usepackage{url}            
\usepackage{booktabs}       
\usepackage{amsfonts}       
\usepackage{nicefrac}       
\usepackage{microtype}      
\usepackage{lipsum}		
\usepackage{graphicx}
\usepackage{natbib}
\usepackage{doi}
\usepackage{amssymb,amsmath}
\usepackage{adjustbox}
\usepackage{rotating}
\usepackage{xcolor,paralist,hyperref,titlesec,fancyhdr,etoolbox}
\usepackage{url}
\usepackage{comment}
\usepackage{tablefootnote}
\usepackage{threeparttable}
\usepackage{enumitem}
\usepackage{bm}
\usepackage{moreverb}

  \title{\bf Optimizing Efficiency and Convergence in MMRM: Practical Considerations for Longitudinal Data Analysis}
  \author{Dylan Clarke\thanks{Correspondence: Dylan Clarke, Department of Statistics, Purdue University, West Lafayette, IN 47907, USA. Email: clarke55@purdue.edu}\hspace{.2cm}\\
    Department of Statistics \\
    Purdue University\\
    West Lafayette, IN 47907, USA \\
    clarke55@purdue.edu
\And
    Jeremiah Jones \\
    Eli Lilly \& Company \\
    Indianapolis, IN 46285, USA \\
    jeremiah.jones@lilly.com
\And
    Yongming Qu \\
    Eli Lilly \& Company\\
    Indianapolis, IN 46285, USA \\
    qu\_yongming@lilly.com}

    \renewcommand{\shorttitle}{\textit{Optimizing Efficiency and Convergence in MMRM}}

    \hypersetup{
pdftitle={Optimizing Efficiency and Convergence in MMRM: Practical Considerations for Longitudinal Data Analysis},
pdfsubject={stat.ME},
pdfauthor={Dylan Clarke, Jeremiah Jones, Yongming Qu},
pdfkeywords={Clinical Trials, Longitudinal Data Analysis, Mixed Models for Repeated Measures},
}

\begin{document}

\maketitle

\begin{abstract}
   Mixed models for repeated measures (MMRM) are a popular method for analyzing longitudinal data in clinical trials. However, practical challenges, such as small sample sizes, large numbers of time points, and selection of variance-covariance structure for within-subject errors, often present barriers to model convergence and valid inference. This article evaluates different options for using MMRM in different scenarios through extensive simulation studies and an application on diabetes trial data. We demonstrate that the empirical bias-reduced coefficient covariance adjustment with the heterogeneous autoregressive covariance structure yields near-nominal coverage with high convergence rates for moderate and large sample designs. For small sample sizes, the simple model with baseline covariates and treatment by time point interaction achieves good efficiency and high probability of convergence. Based on these results, we provide practitioners with actionable guidance for applying MMRM to clinical trial data. 
\end{abstract}


\section{Introduction}
\label{s:intro}

Longitudinal outcomes are commonly collected in clinical trials. Hypothetical strategy is one commonly used strategy to handle intercurrent events in defining estimands in the ICH E9(R1) Addendum \citep{guideline2017addendum}. Missing data almost always occur either due to patient dropouts from the study or censorship due to intercurrent events despite missing data prevention efforts. The mixed model for repeated measures (MMRM) is a well established statistical method for analyzing longitudinal data with missing values under the missing at random (MAR) assumption \citep{gueorguieva2004move,mallinckrod2008recommendations}. The estimand using the hypothetical strategy to handle intercurrent events and the corresponding statistical analysis using MMRM under the MAR assumption provides estimates for the treatment effect if all participants would have adhered to the assigned treatments. Consequently, results based on MMRM have been widely used in articles reporting clinical trial results, e.g., \cite{rosenstock2025weekly}.

Despite well established theory and the popularity, the MMRM approach has been applied in an inconsistent way to clinical trials. This is because there is flexibility in the MMRM options, including the fixed effects, the variance-covariance structure, and the variance estimation for the fixed effects. These options may impact the computation time, model convergence, statistical efficiency of the estimators, and the validity of statistical inferences.
For example, \cite{Gosho2018} and \cite{wang2024improving} suggest MMRM with more fixed effects and variance-covariance parameters could lead to efficiency gain at the cost of poor model convergence for small sample sizes and increased computation time for large sample sizes. 

In this article, we examine the choice of MMRM options in large and small samples through simulation. The performance is evaluated in terms of coverage, convergence rate, bias, and efficiency. A recommendation will be provided based on the simulation results. An application to diabetes trial data is included to proof the results and examine the practicality of the methods.

\section{Methods}
\label{s:methods}

Let $\bm Y_i$ represent the vector of post-baseline longitudinal outcomes for $K$ fixed time points for subject $i \in 1,...,n$. The outcome vector $Y_i$ is modeled as \[
  \bm{Y}_i = \bm{\mu}_{it}(X_i;\bm{\beta}) + \bm{\epsilon}_i,~ \bm{\epsilon}_i \sim N(\bm{0}, {V_i}),
  \label{eq:mmrm-model}
\]
where $\bm{\mu}_{it}$ is a function of baseline covariates $X_i$ (which may include the baseline observation of the outcomes $Y_{i0}$) and depends on a parameter $\bm{\beta}$ which must be estimated. The variable $\bm \epsilon_i$ represents the error term with subject time-point-by-time-point covariance matrix $V_i$.

The vector $\bm R_i \in \{0,1\}^K$ is used to represent the missingness for $K$ time points with 0 for missing value and 1 for observed data. When missingness does not depend on any observed or unobserved value (MCAR) or only depends on observed values (MAR), it can be considered \emph{ignorable} \citep{rubin1976inference}. Valid inference can be made using only the observed values under ignorable missingness. We further consider that each subject had the potential to receive any one of $J+1$ treatments.  Let $j_i \in \{0,1,...,J: J\ge 1\}$ represent the treatment assignment of subject $i$, with $j=0$ representing comparator treatment.

\begin{figure}
    \centering
    \resizebox{\textwidth}{!}{
    \includegraphics[width=0.5\linewidth]{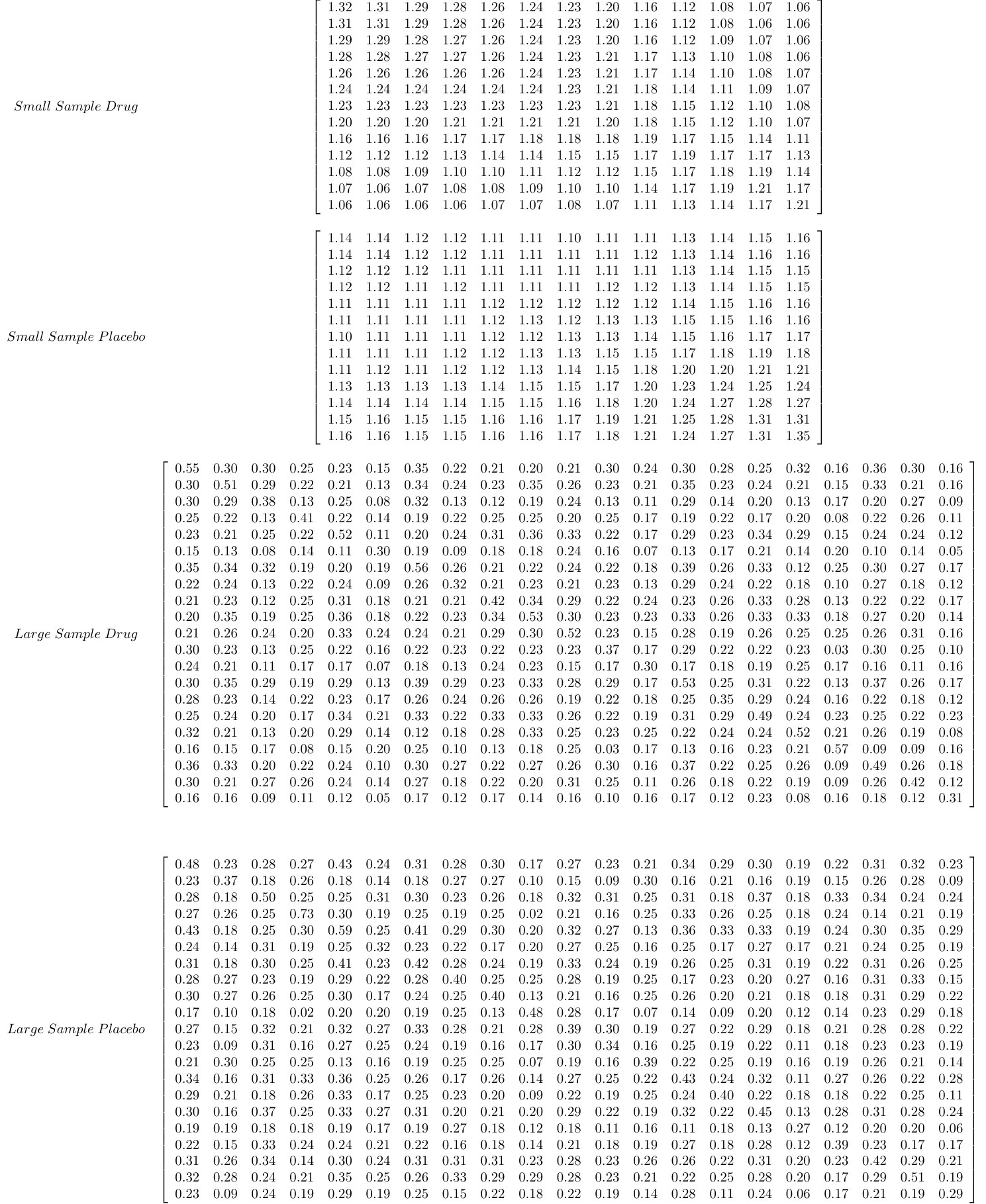}}
    \caption{Variance-covariance matrices used in simulations.}
    \label{fig:cov}
\end{figure}

We mimicked the stratified randomization commonly used in clinical trials by considering $J$ experimental treatment groups and 1 control group in the simulation with $C$ stratification factors. Each stratification factor has 2 levels, leading to $2^C$ randomization strata. The stratified randomization was implemented using the \textit{randomizr} package in R \citep{lim2019randomization,randomizr2023}. For each treatment, data were generated independently from a multivariate normal distribution with distinct variance-covariance structure for different treatment groups. Mean and variance-covariance matrix parameters in the distribution were based on empirical values from a randomized trial evaluating an anti-diabetes treatment. The mean response for subject $i$ at time point $t$ ($t=0$ for baseline) was modeled as \[
    \mu^*_{it} = \beta^*_{0} + \beta^*_{j_it} + \left( \sum_{l=1}^C{2^{l-1}S_{il}} \right) \beta^*_{Xj_it}, \quad t=0, 1, \ldots, K,
\]
where $S_{il}$ is an indicator for an arbitrary level of stratification factor $l$ for subject $i$. The error variance-covariance matrix for each treatment group is given in Figure \ref{fig:cov} where the first column represents the baseline value. 

Patients' dropout was generated based on the observed values.
The probability of dropout at time point $t$ was modeled as 
\[
    \mbox{logit}(P(R_{it} = 1 \mid R_{it-1} = 1)) = \mbox{logit}\left(\frac{1 - p_{t}^*}{1- p_{t-1}^*}\right) - \gamma_{tj} r_{ijt},
\] for $t > 0$ where $p_{t}^*$ is calibrated to control the probability of dropout at or before time point $t$. The quantity $r_{ijt}$ satisfies $r_{ij1}=Y_{i0},$ and for $t>1,~ r_{ijt}=Y_{it-1}-\tilde{Y}_{it-1}$ for $\tilde{Y}_{it-1}$, the predicted value of ${Y}_{it-1}$ from a regression of $Y_{it-1}$ fit in treatment group $j.$ The initial linear coefficient is $\gamma_{00} = 0.14$ and $\gamma_{0j} = 0.12$ when $j > 0$, and for $t>0$, $\gamma_{t0} = 0.7$ and $\gamma_{tj} = 0.5$ when $j>0$.

Generated data were analyzed using MMRM. The outcome variable was change from baseline, and the estimand of interest was the mean difference between a treatment group $j > 1$ and the comparator group $j = 0$. Three MMRM fixed-effects models of increasing complexity are considered for estimation of 
the fixed effect:
\begin{eqnarray}
    \bm{\mu}_{it}(X_i;\bm{\beta}^{(1)}) =  {\beta_{0}^{(1)}} + \bm{X}_{i}\bm{\beta_{X}^{(1)}} + {\beta_{j_it}^{(1)}} \label{eq:simple}\\
\bm{\mu}_{it}(X_i;\bm{\beta}^{(2)}) =  {\beta_0^{(2)}} + \bm{X}_{i}\bm{\beta}_{\bm{X}t}^{(2)} + {\beta_{j_it}^{(2)}} \label{eq:2way} \\
\bm{\mu}_{it}(X_i;\bm{\beta}^{(3)}) = {\beta_0^{(3)}} + \bm{X}_{i}\bm{\beta}_{\bm{X}t}^{(3)} + {\beta_{j_it}^{(3)}} + \bm{X_{it}}\bm{\beta}_{\bm{X}j_it}^{(3)} ,
\label{eq:3way}
\end{eqnarray} where $X$ contains $Y_{i0}$ and binary encoding for randomization strata as covariates. For example, if $C=2$, then $X_i=(Y_{i0},S_{i1},S_{i2})$. These models (\ref{eq:simple}), (\ref{eq:2way}), and (\ref{eq:3way}) are referred to as the simple, two-way, and three-way fixed-effect fit \citep{wang2024improving}. When fitting simple and two-way fixed-effect models, the same variance-covariance matrix was modeled for the error terms across treatment groups (homogeneous variance); when fitting the three-way fixed effect model, the error term of each treatment group used a different variance-covariance matrix (heterogeneous variance). 

The covariance structures and coefficient covariance adjustment methods considered in the analysis were selected from the list of available methods in the \textit{mmrm} package in R \citep{mmrm}. The unstructured covariance structure (UN), heterogeneous Toeplitz covariance structure (TOEPH), heterogeneous autoregressive-1 covariance structure (AR1H), homogeneous compound symmetry (CS), and heterogeneous compound symmetry (CSH) covariance structures are considered alongside the asymptotic, empirical, empirical bias-reduced (EBR), empirical jackknife (EJK), and Kenward-Roger (KR) coefficient covariance adjustments \citep{kauermann2001note,mancl2001covariance,kenward1997small,bell2002bias}. Between-within degree of freedom approximation for the $t$-distribution are used in the formation of confidence intervals, except when using KR coefficient covariance adjustments \citep{mmrm}. 

Analysis was conducted by fitting models in the \textit{mmrm} 0.3.15 package in R 4.3.2 followed by extracting least squares empirical marginal means using the \textit{emmeans} 1.11.1 package \citep{mmrm,emmeans,r2024r}. If either package returned an estimation error, i.e. numerical estimation failed to converge, the model would be deemed to have not converged and no further analysis would be preformed. Otherwise, the estimated mean difference (estimate) would be reported for each experimental treatment group versus control group alongside an associated bias, standard error, and 95\% confidence interval (CI). The standard error and the 95\% CI were estimated by considering the randomness in $X$. The time to convergence of the MMRM model (time) would also be returned. Various metrics were produced and summarized in these simulations. The average bias of the estimate (bias), the coverage rate (coverage rate) of the 95\% CI, mean squared error (MSE), empirical standard deviation of estimate (ESD), average standard error (SE), and ratio of SE to ESD (ratio SE) would be calculated for each treatment group along with model convergence rate (CR) across 1000 simulated data sets. 

Simulation is divided into two categories: a large sample category, likely representative of late stage clinical trials where estimation issues are due to the large number of subjects and time points, and a small sample category, likely representative of early stage clinical trials where estimation issues are due to small samples sizes. For the large sample case, simulations are conducted under a single setting with $n = 800$, $J = 1$, $C = 3$, and $K = 20$.
For the small sample case, simulations are conducted under several settings using $n \in \{12,18,24,30,36,42,48,54,60,66,72,78,84,90,150\}$, $J = 2$, $C \in \{2,3,4\}$, and $K \in \{4,6,8,10,12\}$. 

An additional setting where the true error variance depends on the baseline value is tested for the large samples (heteroskedasticity). For the small sample category, a comparison of algorithms for fitting an MMRM model is conducted. Models with various options were compared with traditional data-adaptive techniques in terms of the previously defined metrics in order to justify our suggested algorithm. Finally, further justification of the final algorithm is presented in terms of these metrics.

\section{Results}
\label{s:results}

\subsection{Large Sample Results}
Table \ref{t:large} shows the simulation results for the setting with a large sample size. Bias is negligible for all option combinations, so it is excluded from the table.
Under a correctly specified model, represented by No Heteroskedasticity in Table \ref{t:large}, the speed of convergence for the model fit separately for each treatment by allowing for different fixed effects and error variance covariance matrix is faster than the two-way fit outside of the UN covariance structure, whereas the Ratio SE, MSE, and Coverage Rate are all similar. In all covariance structures tested, the use of EBR brings the ratio SE closer to the desired value of one.  The use of the two-way model offers little benefit as it is slower than the more robust three-way model with EBR and a simple covariance structure.  Furthermore, the MSE and coverage rates are comparable under the UN and AR1H covariance structures, but the AR1H fits in less time. Additionally, Table \ref{t:large} exhibits the effect of heteroskedasticity; an inflated MSE is observed, but the ratio SE and coverage rate are similar to the correctly specified case. The benefit of the use of EBR and the AR1H covariance matrix are still present in this case as well.

\begin{table}[ht]
\centering
\resizebox{\textwidth}{!}{
\begin{threeparttable}
\caption{Convergence time, Ratio SE, MSE, and coverage rate for large sample simulations\tnote{1}.} 
\begin{tabular}{lllrrrrrrrr}
 \hline
   & & & \multicolumn{4}{l}{No Heteroskedasticity} & \multicolumn{4}{l}{Heteroskedasticity} \\ \hline
 Fit & Covariance Structure & Covariance Adjustment & Time & Ratio SE & MSE & Coverage Rate & Time & Ratio SE & MSE & Coverage Rate \\ 
  \hline
   
  Simple & AR1H & Asymptotic & 0.13 & 1.04 & 0.0014 & 95.80 & 0.12 & 1.00 & 0.0025 & 94.80 \\ 
   &  & Empirical-Bias-Reduced & 0.34 & 1.01 & 0.0014 & 94.90 & 0.34 & 0.99 & 0.0025 & 94.50 \\ 
   & CS & Asymptotic & 0.11 & 1.14 & 0.0013 & 97.70 & 0.14 & 1.28 & 0.0024 & 98.70 \\ 
   &  & Empirical-Bias-Reduced & 0.32 & 1.01 & 0.0013 & 95.20 & 0.35 & 1.00 & 0.0024 & 94.50 \\ 
   & CSH & Asymptotic & 0.17 & 1.03 & 0.0014 & 95.80 & 0.14 & 1.00 & 0.0024 & 94.60 \\ 
   &  & Empirical-Bias-Reduced & 0.38 & 1.01 & 0.0014 & 95.10 & 0.35 & 1.00 & 0.0024 & 94.50 \\ 
   & TOEPH & Asymptotic & 0.26 & 1.05 & 0.0013 & 95.70 & 0.14 & 1.00 & 0.0024 & 94.90 \\ 
   &  & Empirical-Bias-Reduced & 0.48 & 1.01 & 0.0013 & 95.20 & 0.36 & 0.99 & 0.0024 & 94.30 \\ 
   & US & Asymptotic & 0.21 & 1.02 & 0.0013 & 95.20 & 0.21 & 1.00 & 0.0024 & 95.00 \\ 
   &  & Empirical-Bias-Reduced & 0.45 & 1.02 & 0.0013 & 95.20 & 0.45 & 0.99 & 0.0024 & 95.00 \\ 
Two-way & AR1H & Asymptotic & 0.51 & 1.04 & 0.0014 & 95.60 & 0.50 & 1.00 & 0.0025 & 94.70 \\ 
   &  & Empirical-Bias-Reduced & 1.55 & 1.01 & 0.0014 & 94.80 & 1.60 & 0.99 & 0.0025 & 94.50 \\ 
   & CS & Asymptotic & 0.48 & 1.13 & 0.0014 & 97.80 & 0.63 & 1.29 & 0.0024 & 98.60 \\ 
   &  & Empirical-Bias-Reduced & 1.47 & 1.01 & 0.0014 & 95.10 & 1.66 & 1.00 & 0.0024 & 94.40 \\ 
   & CSH & Asymptotic & 0.71 & 1.03 & 0.0014 & 95.90 & 0.58 & 1.00 & 0.0024 & 94.60 \\ 
   &  & Empirical-Bias-Reduced & 1.73 & 1.01 & 0.0014 & 95.40 & 1.64 & 1.00 & 0.0024 & 94.30 \\ 
   & TOEPH & Asymptotic & 1.03 & 1.05 & 0.0013 & 96.00 & 0.54 & 1.00 & 0.0024 & 94.80 \\ 
   &  & Empirical-Bias-Reduced & 2.10 & 1.01 & 0.0013 & 95.20 & 1.74 & 1.00 & 0.0024 & 94.40 \\ 
   & US & Asymptotic & 0.72 & 1.02 & 0.0013 & 95.10 & 0.70 & 1.00 & 0.0024 & 94.90 \\ 
   &  & Empirical-Bias-Reduced & 1.90 & 1.02 & 0.0013 & 95.10 & 2.03 & 1.00 & 0.0024 & 94.70 \\ 
  Three-way & AR1H & Asymptotic & 0.46 & 1.03 & 0.0014 & 95.20 & 0.47 & 1.00 & 0.0025 & 95.10 \\ 
   &  & Empirical-Bias-Reduced & 0.85 & 1.00 & 0.0014 & 94.30 & 0.86 & 0.99 & 0.0025 & 95.00 \\ 
   & CS & Asymptotic & 0.43 & 1.13 & 0.0014 & 97.60 & 0.50 & 1.29 & 0.0024 & 98.80 \\ 
   &  & Empirical-Bias-Reduced & 0.81 & 1.00 & 0.0014 & 95.00 & 0.89 & 1.00 & 0.0024 & 94.40 \\ 
   & CSH & Asymptotic & 0.69 & 1.02 & 0.0014 & 95.90 & 0.57 & 1.00 & 0.0024 & 94.90 \\ 
   &  & Empirical-Bias-Reduced & 1.07 & 1.00 & 0.0014 & 95.10 & 0.96 & 1.00 & 0.0024 & 94.70 \\ 
   & TOEPH & Asymptotic & 0.93 & 1.03 & 0.0014 & 95.80 & 0.53 & 1.00 & 0.0024 & 95.10 \\ 
   &  & Empirical-Bias-Reduced & 1.36 & 1.00 & 0.0014 & 95.70 & 0.95 & 1.00 & 0.0024 & 94.90 \\ 
   & US & Asymptotic & 6.64 & 1.10 & 0.0014 & 96.70 & 0.76 & 1.00 & 0.0024 & 94.80 \\ 
   &  & Empirical-Bias-Reduced & 7.25 & 1.10 & 0.0014 & 96.70 & 1.20 & 1.00 & 0.0024 & 94.80 \\
   \hline
\end{tabular}
\begin{tablenotes}\footnotesize
\item[1] 
Abbreviations: No Heteroskedasticity, correctly specified model; Heteroskedasticity, error variance depends on the baseline value; Time, time to convergence in minutes; Ratio SE, ratio of average standard error to the empirical standard deviation of the estimate; MSE, the mean squared error; Coverage Rate, coverage rate of 95\% confidence interval adjusted for randomness of $X$.
\end{tablenotes}
\end{threeparttable}}
\label{t:large}
\end{table}

\subsection{Small Sample Results}

Table \ref{t:small} summarizes simulation results for the small sample cases with the three-way fit under settings $n = 48$, $C = 3$, and $K=8$. Bias is negligible for all option combinations, so it is again excluded from the table. The results in this table are generally consistent for all settings. Consistent with the results from \cite{mancl2001covariance}, the empirical covariance estimation provides an underestimate of the emprical standard error. KR produces inconsistent coverage rates across covariance structures and frequently underestimates the empirical variance. EJK frequently results in drastic overestimates of the empirical variance, as shown in Ratio SE from Table \ref{t:small}, and inflated coverage rates. In contrast, EBR produces a consistent level of coverage close to the nominal level across covariance structures. EBR combined with the AR1H covariance structure produced high levels of convergence at small sample sizes while maintaining estimated variances in agreement with the empirical variances, small mean squared errors, and coverage rates near nominal.

\begin{figure}
    \centering
        \resizebox{\textwidth}{!}{
    \includegraphics[width=0.5\linewidth]{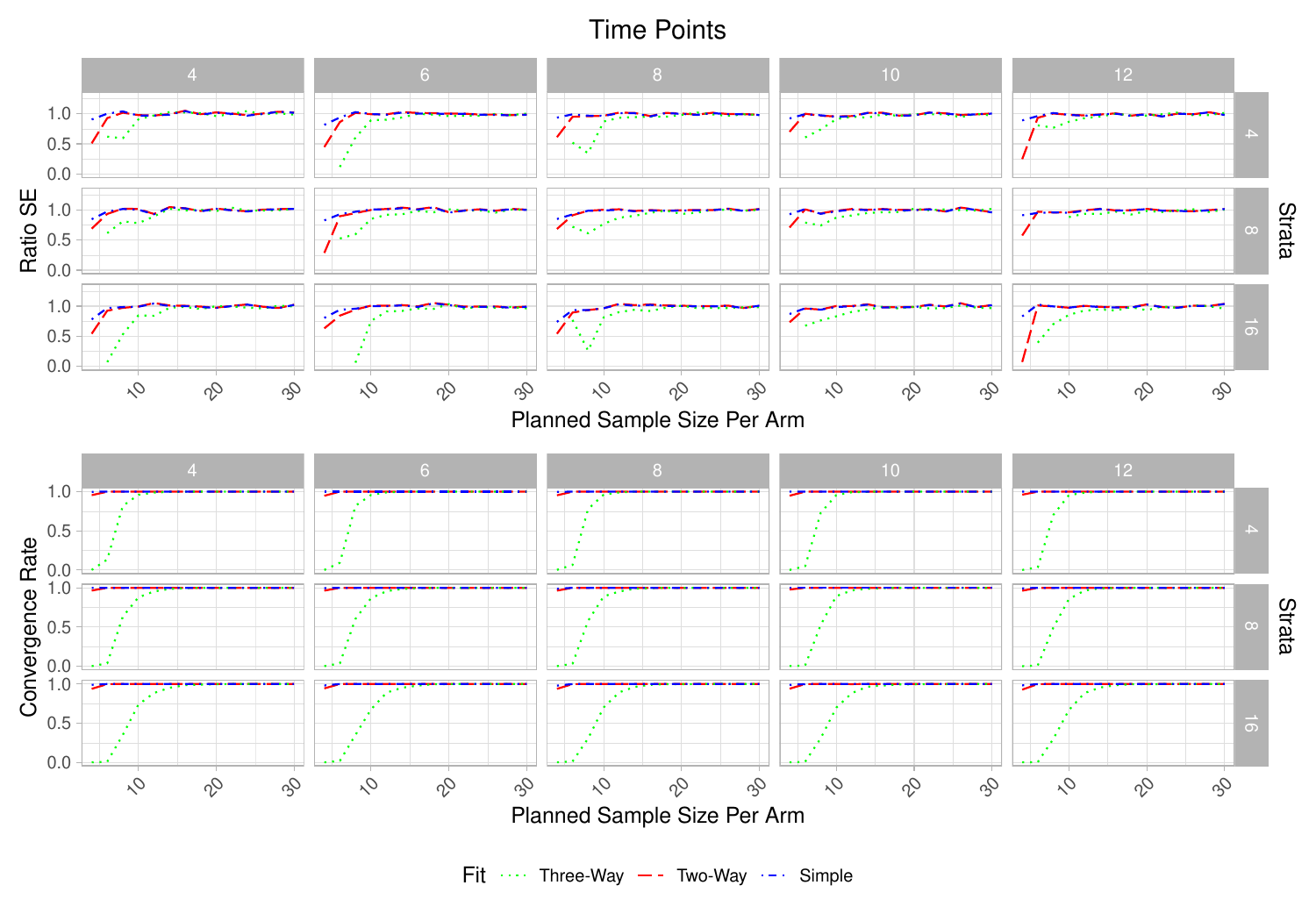}}
    \caption{Difference between models by strata count, number of time points, and smallest treatment group by subjects not withdrawn when using the empirical bias-reduced coefficient covariance adjustment and heterogeneous autoregressive error variance-covariance structure.}
    \label{fig:modeldifferences}
    \end{figure}

\begin{figure}
    \centering
 \hspace{24cm}
        \resizebox{\textwidth}{!}{
    \includegraphics[width=0.5\linewidth]{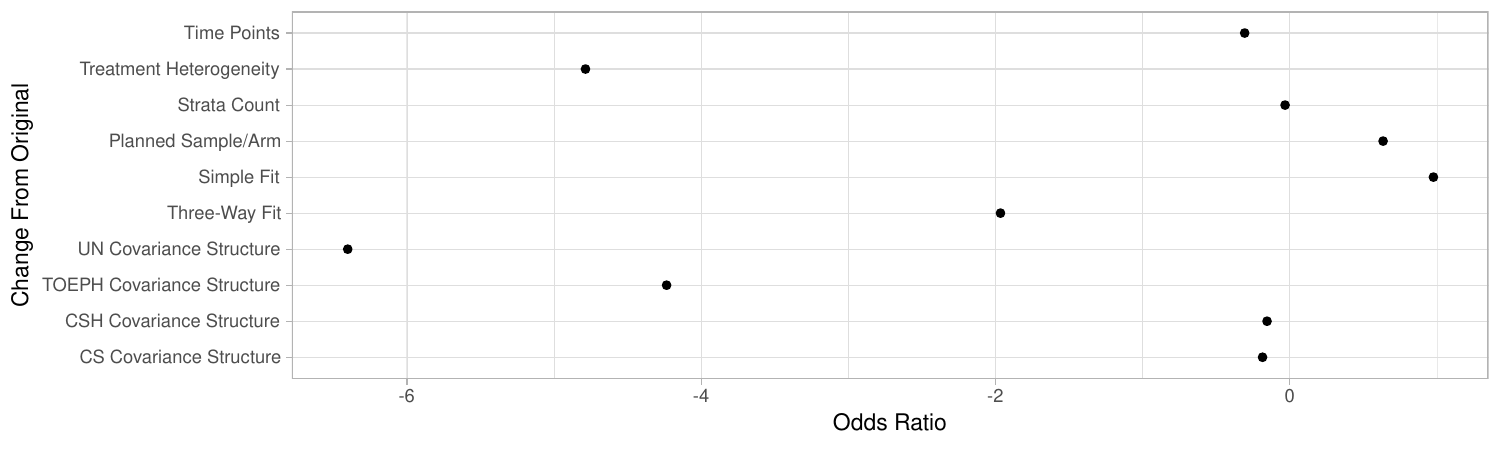}}
    \caption{Logistic regression on convergence of individual models expressed as changes from two-way model with AR1H. The number of time points, strata count, and planned sample are treated as continuous variables, and the rest are discrete variables.}
    \label{fig:convplotdots}
\end{figure}

Figure \ref{fig:modeldifferences} shows the ratio SE and convergence rate against the number of strata, the number of time points, and sample size. As expected, the convergence rate of the three-way model is lower than the two simpler models. The convergence rate is approximately constant across the number of time points when strata is held constant. The SE tends to be more extremely underestimated and unstable in the three-way model for 10 or fewer subjects per treatment group. The number of strata has a minimal impact on the convergence of the model. In extremely small samples, the two- and three-way models underestimate the ESD compared to the simple model, which obtained a Ratio SE close to 1 with fewer samples.
Additional analyses show that options with large impact on the number of parameters (heterogeneous treatment, UN, and TOEPH variance-covariance structure, and the three-way model) reduce the probability of model convergence (Figure \ref{fig:convplotdots}). 

\begin{figure}
    \centering
    \resizebox{\textwidth}{!}{
    \includegraphics[width=0.5\linewidth]{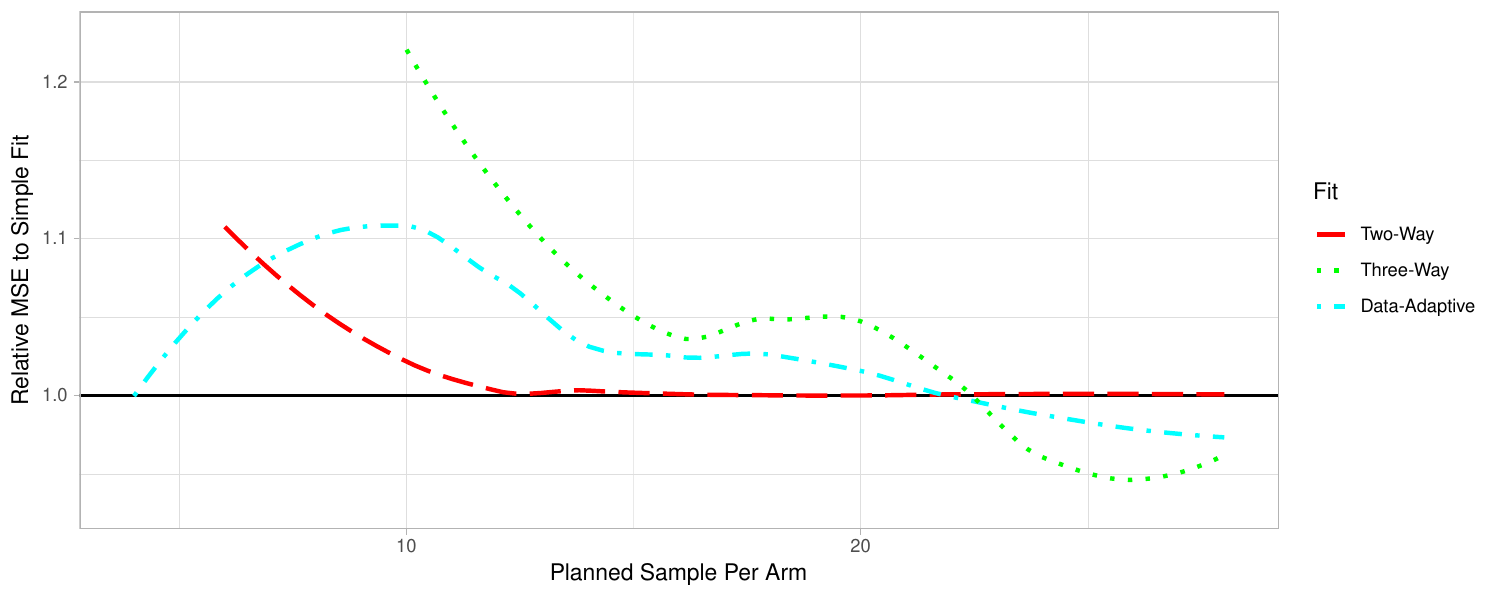}}
    \caption{Relative mean squared error compared with simple fit for 12 time points and 3 binary stratification variables. The data-adaptive fit attempts to fit the three-way fit first, followed by the simple fit if the original model fails to converge.}
    \label{fig:process}
\end{figure}

For planned sample sizes per arm less than 25, we compare the results of convergence iteration procedures, where models are fit with reducing complexity until a model converges, with results from using a predetermined model based on the design of the study. Figure \ref{fig:process} shows a comparison of a convergence iterative procedure versus predetermined models. The MSEs of the two-way, three-way, and data-adaptive models (i.e., reducing model complexity until convergence is achieved) were mostly larger than the simple model when the sample sizes are small.

\section{Recommendations}
\label{s:Recommendations}

Based on the results from these simulations, we 
found the use of the three-way fit with an AR1H covariance structure and the EBR covariance estimation resulted in very fast convergence and superior efficiency when the sample size was moderate or large (e.g., $\ge 25$ per treatment group). When the sample size was small (e.g., $< 25$ per treatment group), the simple model ensured high probability of model convergence while also tending to maintain a superior MSE. The data-adaptive approach, which attempts the most complex model and decreases complexity iteratively until convergence is achieved, resulted in poor performance. Therefore, we
recommend the following for planning clinical studies:
\begin{enumerate}[label={\textbf{REC-\arabic*}},leftmargin=*,]
    \item{For sample sizes approximately less than 25 per arm, use a simple model with homogeneous variance across treatment groups. For sample sizes greater than this figure, use the three-way model with heterogeneous variance.}
    \item{The UN error covariance structure is unnecessary when using robust parameter covariance estimates. Fitting an AR1H covariance provides a good balance of computational and statistical efficiency.}
    \item{Asymptotic, Kenward-Roger, and standard Empirical parameter covariance estimators provide poor statistical performance. Empirical-Bias-Reduced estimators are robust across all tested situations and are not sensitive to the choice of error covariance structure.}
\end{enumerate}

\begin{table}[ht]
\centering
\resizebox{\textwidth}{!}{
\begin{threeparttable}
\caption{Ratio SE, MSE, Coverage Rate, and Inflate SE for the first treatment level for the three-way fit with 3 treatment levels, 48 subjects, 8 time points, and 3 stratification variables\tnote{1}.} 
\begin{tabular}{llrrrr}
  \hline
Coefficient Covariance Adjustment & Covariance Structure & CR & Ratio SE & MSE & Coverage Rate\\ 
  \hline
Asymptotic & AR1H & 99.40 & 0.83 & 0.0205 & 92.47\\ 
   & CS & 99.40 & 0.60 & 0.0212 & 79.92 \\ 
   & CSH & 99.40 & 0.97 & 0.0208 & 96.59 \\ 
   & TOEPH & 96.80 & 0.89 & 0.0208 & 93.51 \\ 
   & UN & 62.90 & 0.96 & 0.0208 & 96.20 \\ 
  Empirical & AR1H & 99.40 & 0.76 & 0.0205 & 89.56  \\ 
   & CS & 99.40 & 0.75 & 0.0212 & 88.86  \\ 
   & CSH & 99.40 & 0.77 & 0.0208 & 89.26  \\ 
   & TOEPH & 96.80 & 0.76 & 0.0208 & 89.79  \\ 
   & UN & 62.90 & 0.77 & 0.0208 & 90.33  \\ 
  Empirical-Bias-Reduced & AR1H & 99.40 & 0.95 & 0.0205 & 94.98  \\ 
   & CS & 99.40 & 0.95 & 0.0212 & 94.98  \\ 
   & CSH & 99.40 & 0.96 & 0.0208 & 95.58  \\ 
   & TOEPH & 96.80 & 0.95 & 0.0208 & 94.74  \\ 
   & UN & 62.90 & 0.95 & 0.0208 & 95.09  \\ 
  Empirical-Jackknife & AR1H & 99.40 & 1.60 & 0.0205 & 98.59  \\ 
   & CS & 99.40 & 1.99 & 0.0212 & 98.69  \\ 
   & CSH & 99.40 & 1.43 & 0.0208 & 98.80  \\ 
   & TOEPH & 96.80 & 1.67 & 0.0208 & 98.45  \\ 
   & UN & 62.90 & 1.42 & 0.0208 & 98.26  \\ 
  Kenward-Roger & AR1H & 99.40 & 0.80 & 0.0205 & 90.36  \\ 
   & CS & 99.40 & 0.59 & 0.0212 & 76.10  \\ 
   & CSH & 99.40 & 0.92 & 0.0208 & 95.38  \\ 
   & TOEPH & 96.80 & 0.86 & 0.0208 & 92.06  \\ 
   & UN & 62.90 & 0.83 & 0.0208 & 92.23  \\ 
   \hline
\end{tabular}
\begin{tablenotes}\footnotesize

\item[1] Abbreviations: CR,  convergence rate; Ratio SE, ratio of average standard error to the empirical standard deviation of the estimate; MSE, mean squared error; Coverage Rate, coverage rate of 95\% confidence interval adjusted for randomness of $X$.
\end{tablenotes}
\end{threeparttable}}
\label{t:small}
\end{table}

\section{An Application in Clinical Trials}
\label{s:application}

We applied our recommendations to a randomized, open-label study comparing weekly fixed-dose insulin to daily insulin \citep{rosenstock2025weekly}. The data used for the analysis contained information for 795 unique patients randomized in a 1:1 ratio over the course of 52 weeks, stratified by country of residence, a binary variable for Hemoglobin A1c (HbA1c) at baseline ($< 8\%$, or $\geq 8\%$), Glucagon-Like Peptide-1 (GLP-1) receptor agonist usage at randomization (yes or no). The primary analysis compared the change from baseline in HbA1c at week 52, with a secondary analysis comparing participant-monitored fasting blood glucose (FBG). Throughout this discussion, we focus on the efficacy estimand, which was analyzed using MMRM.

The pre-specified model was a simple model (including the three stratification variables as additive factors) using a UN covariance structure, with a series of simpler models used in case of convergence issues. For the primary analysis, where the continuous baseline value of HbA1c was used, the HbA1c stratum covariate was not included. We compare the pre-specified model (a) with our recommended three-way model using (b) AR1H and (c) CSH, to compare the computational tradeoffs, estimates, and standard errors. The SEs for each model are provided by the EBR estimator based on our simulation results.

For the primary endpoint analysis, total of 8 unique post-baseline timepoints were included, while the FBG analysis included 29 \citep[Fig. 2A and 2C, respectively, in][]{rosenstock2025weekly}.  These analyses showcase two contrasting motivations of the recommendations: the analysis of HbA1c involved only 28 parameters for the pre-specified UN covariance structure, while a similar analysis of FBG required an infeasible 406 parameters for that UN structure. 

To highlight these computational issues, we begin with the FBG analysis. Using the simple model with a UN covariance structure failed to converge with 2 hours of run time and 42 gigabytes of memory. The trial reported results using a simple model with a CS covariance structure--conesequently, we adopt CS for our model (a). Model (a) produced an estimated treatment difference between the weekly insulin and the daily comparator (ETD) of 1.02 with SE of 2.75 in 77 seconds. Model (b) produced an ETD of 2.62 with SE of 2.70 in 206 seconds. Model (c) produced an ETD of 0.46 with SE of 2.82 in 175 seconds. All three models are generally consistent with slightly elevated but statistically insignificant differences in FBG when comparing the weekly insulin with a daily comparator. This is intuitive, as participants in both arms were instructed to titrate insulin to reach a target FBG. Among the three models compared, the recommended model achieved the smallest SE in a computational time under 5 minutes. In practice, the CS model used in (a) would have been selected after failing UN.

Turning to the primary analysis of HbA1c at week 52, the smaller number of timepoints allows UN for Model (a). The updated model (a) resulted in an ETD of -0.037 with a SE of 0.080 in 7 seconds. Model (b) produced an ETD of -0.082 with SE of 0.071 in 4 seconds. Model (c) produced an ETD of -0.059 with SE of 0.073 also in 4 seconds.  Once again, the recommended model (b) provided the smallest SE among the three models and achieved the lowest time to fit compared to model (a).

\section{Summary and Discussion}
\label{s:discussion}

MMRM is often used in analysis of longitudinal data with missing outcomes under the MAR assumption. There is a need for a consistent procedure to apply MMRM to clinical trials. 
We compared different MMRM options through simulation studies for small and large samples. We found that models with fewer variance-covariance parameters in conjunction with the EBR variance adjustment method could perform just as well in large samples and better in small samples than models with more variance-covariance parameters.

Hence, we recommend the use of the three-way model (REC-1) with AR1H covariance (REC-2) and EBR coefficient covariance estimator (REC-3) to achieve statistical and computational efficiency alongside robustness to modeling assumptions when the planned sample size per arm is 25 or greater. If the sample size is less than 25 per treatment group, the simple model (REC-1) should be used alongside similar covariance procedures (REC-2 and REC-3).

An important limitation of this result is that simulated data is simulated using a normal distribution for the outcome and a true mean model that closely follows the form of the mean models tested in this paper. Data from real clinical trials are likely to be much more complicated and no ground truth distribution can be known. However, since the simulation setting was close to a real clinical trial data and is robust in term of number of strata, number of treatment groups, and number of post-baseline time points, we expect the conclusion drawn from this article should be applied to future trial data. Future research could extend these results to different data generating processes, such as heavy tail distributions or higher levels of missingness. Additionally, it could be of interest to retest the simulations under different criteria for ``convergence'' since the more complicated models were reporting convergence but performing poorly in extremely small sample sizes.

In summary, through extensive simulation, this article provides a practical recommendation for MMRM options in clinical trials to achieve optimal efficiency, convergence, and computation speed. The recommendation should be considered in longitudinal data analysis in future clinical trials. 

\section*{Data Availability Statement}

The authors elect not to share data.

\section*{Funding Statement}

This work was supported by Eli Lilly and Company.

\section*{Conflicts of Interest}

The authors declare no conflicts of interest.

\section*{Ethical Statement}

Any clinical trial data referenced in this manuscript were previously collected under protocols approved by relevant institutional review. No new trial data was collected.

\bibliographystyle{apalike}
\bibliography{references}

\end{document}